\documentclass{article}
\usepackage{amssymb}
\usepackage[centertags]{amsmath}
\usepackage{graphicx}

\begin{document}

\title{The essence of the homotopy analysis method}

\author{ Cheng-shi Liu \\Department of Mathematics\\Daqing Petroleum Institute\\Daqing 163318, China
\\Email: chengshiliu-68@126.com\\Tel:+86-459-6503476}

 \maketitle

\begin{abstract}
The generalized Taylor expansion including a secret auxiliary
parameter $h$ which can control and adjust the convergence region of
the series is the foundation of the homotopy analysis method
proposed by Liao. The secret of $h$ can't be understood  in the
frame of the homotopy analysis method. This is a serious shortcoming
of Liao's method. We solve the problem. Through a detailed study of
a simple example, we show that the generalized Taylor expansion is
just the usual Taylor's expansion at different point $t_1$. We prove
that there is a relationship between $h$ and $t_1$, which reveals
the meaning of $h$ and the essence of the homotopy analysis method.
As an important example, we study the series solution of the Blasius
equation. Using the series expansion method at different points, we
obtain the same result with liao's solution given by the homotopy
analysis method.


\end{abstract}

\section{Introduction}

In a series of papers[1-12], Liao developed and applied the homotopy
analysis method to deal with  a lot of nonlinear problems. In 2004,
Liao published the book[13] in which he summarized the basic ideas
of the homotopy analysis method and gave the details of his approach
both in the theory and on a large number of practical examples. The
key of his method is the generalized Taylor expansion. We next
describe this result. For a given nonlinear differential equation
with initial conditions at the point $t_0$, we can obtain the power
series solution
\begin{equation}
f(t)=\sum_{n=0}^{+\infty}\frac{f^{(n)}(t_0)}{n!}(t-t_0)^n,
\end{equation}
with the convergence region $|t-t_0|<\rho_0$. Through introducing an
approaching function, Liao gives the following so-called generalized
Taylor series solution to the nonlinear equation considered,
\begin{equation}
f(t)=\lim_{m\rightarrow
\infty}\sum_{n=0}^{m}\mu_{m,n}(h)\frac{f^{(n)}(t_0)}{n!}(t-t_0)^n,
\end{equation}
where the approaching function $\mu_{m,n}(h)$ satisfies
$\lim_{m\rightarrow \infty}\mu_{m,n}(h)=1$ when $n\geq 1$. Liao
points out that the generalized Taylor series provides a way to
control and adjust the convergence region through an auxiliary
parameter $h$ such that the homotopy analysis method is particularly
suitable for problems with strong nonlinearity[3]. However, the
mathematical meaning of the parameter $h$ is still unknown. This is
a serious shortcoming of the homotopy  analysis method.

In the present paper, we point out that the so-called generalized
Taylor series at the initial point $t_0$ is just the usual Taylor
expansion of $f(t)$ at another point. Our results give the meaning
of $h$ in the generalized Taylor expansion, and also uncover the
essence of the homotopy analysis method. We study two examples in
detail.  Especially, using the series expansion method to solve the
Blasius equation, we give the same result with the Liao's
generalized Taylor series solution by the homotopy analysis method.

\section{The detailed analysis of a simple example}

In the book [13], Liao had studied  a simple example
$f(t)=\frac{1}{1+t}$ to illustrate the generalized Taylor series. He
gave the following generalized Taylor expansion,
\begin{equation}
f(t)=\lim_{m\rightarrow
\infty}\sum_{n=0}^{m}\mu_0^{m,n}(h)[(-1)^nt^n],
\end{equation}
with the convergence region
\begin{equation}
-1<t<\frac{2}{|h|}-1, (-2<h<0),
\end{equation}
 where
\begin{equation}
\mu_0^{m,n}(h)=(-h)^{n+1}\sum_{k=0}^{m-n} \binom{n+k}{n}(1+h)^k.
\end{equation}
 Liao concludes that the series (3) has enlarged
the convergence region of the Taylor series of $f$ at the point
$t=0$ and calls it the generalized Taylor series.

We next analyze the essence of the generalized Taylor expansion in
detail. We prove that the above generalized Taylor series (3) is
just the usual Taylor expansion at the point $t_0=-1-\frac{1}{h}$.
Indeed, the Taylor series of $f$ at the point $t=0$ is
\begin{equation}
f(t)=1-t+\cdots+(-1)^nt^n+\cdots,
\end{equation}
with the convergence region $|t|<1$. The Taylor series at the point
$t_0$ is
\begin{equation}
f(t)=\frac{1}{1+t_0}\{1-\frac{t-t_0}{1+t_0}+\cdots+(-1)^n(\frac{t-t_0}{1+t_0})^n+\cdots\},
\end{equation}
with the convergence region $|t-t_0|<|1+t_0|$. If we take
$t_0=-1-\frac{1}{h}$,  the above expression (7) becomes
\begin{eqnarray}
f(t)=-h\{1+h(t+1+\frac{1}{h})+\cdots+(-h)^n(t+1+\frac{1}{h})^n+\cdots\}\cr
=-h\sum_{n=0}^{+\infty}(-1)^n\sum_{k=0}^n
\binom{n}{k}t^k(1+h)^{n-k}(-1)^nh^{k}\cr
=\sum_{k=0}^{+\infty}(-1)^kt^k\sum_{n=k}^{+\infty}
\binom{n}{k}(1+h)^{n-k}(-h)^{k+1},
\end{eqnarray}
which is just the generalized Taylor series (3).

When $t_0>-\frac{1}{2}$, that is, $-2<h<0$, we have
\begin{equation}
\lim_{m\rightarrow \infty}\mu_{m,n}(h)=\lim_{m\rightarrow
\infty}\frac{1}{(1+t_0)^{k+1}}\sum_{n=k}^m
\binom{n}{k}(-t_0)^{n-k}=1.
\end{equation}
Correspondingly, the convergence region $|t-t_0|<|1+t_0|$ is just
(4).

When $-1\leq t_0<-\frac{1}{2}$ or $t_0<-1$, we have
\begin{equation}
\lim_{m\rightarrow \infty}\mu_{m,n}(h)=\lim_{m\rightarrow
\infty}\frac{1}{(1+t_0)^{k+1}}\sum_{n=k}^m
\binom{n}{k}(-t_0)^{n-k}=\infty.
\end{equation}

Therefore, when $t_0>-\frac{1}{2}$ and $-1<t<1+2t_0$, that is,
$-2<h<0$ and $-1<t<-1-\frac{2}{h}=\frac{2}{|h|}-1$,  the right side
of expression (3) is just the Taylor series of $f$ at the point
$t_0=-1-\frac{1}{h}$. From our discussion and computation, it is
easy to see that as the Taylor series of $f$ at the point $t_0$ the
convergence region of the series (3) depends on the point $t_0$.
When $t_0\rightarrow +\infty$, that is, $h\rightarrow 0$, the
convergence region $(-1, 2t_0+1)$ of the series (3) becomes $(-1,
+\infty)$ naturally. In other words, $t_0$ can  be used to control
and adjust the convergence region.

Our result gives the meaning of the auxiliary parameter $h$, and
hence uncovers the essence of the generalized Taylor expansion as
the kernel of the homotopy analysis method. In general, we have the
following conclusion: the generalized Taylor series at the initial
point $t_0$ is only the usual Taylor expansion of $f(t)$ at another
point $t_1=(\eta_0-t_0)(1+\frac{1}{h})+t_0$, where the real number
$\eta_0$ satisfies $|\eta_0-t_1|=|\xi_0-t_1|$, and $\xi_0$ is the
nearest singular point of $f(t)$ from $t_1$. Indeed, let $\xi_k$ (
$k=1,2,\cdots$)  be all singular points of a function $f$ which is
an analytic function at the point $t_1$ and the convergence radius
be $\rho_0=\inf(|t_1-\xi_k|, k=1,2,\cdots)$. When $t_0$ belongs to
the convergence region of the Taylor expansion of $f$ at the point
$t_1$, we can represent the Taylor series at the point $t_1$ by the
Taylor series at the point $t_0$ as follows,
\begin{eqnarray}
f(t)=\sum_{n=0}^{+\infty}\frac{f^{(n)}(t_1)}{n!}(t-t_1)^n
=\sum_{n=0}^{+\infty}\sum_{k=0}^{n}\frac{f^{(n)}(t_1)}{n!}\binom{n}{k}(t-t_0)^{k}(t_0-t_1)^{n-k}\cr
=\sum_{n=0}^{+\infty}\sum_{k=n}^{+\infty}\frac{f^{(k)}(t_1)}{k!}\binom{k}{n}(t_0-t_1)^{k-n}(t-t_0)^{n}\cr
=\lim_{m\rightarrow
\infty}\sum_{n=0}^{m}\mu_{m,n}(f;t_0,t_1)\frac{f^{(n)}(t_0)}{n!}(t-t_0)^n,
\end{eqnarray}
where the approaching function
\begin{equation}
\mu_{m,n}(f;t_0,t_1)=\{\sum_{k=n}^{m}\frac{f^{(k)}(t_1)}{k!}\binom{k}{n}(t_0-t_1)^{k-n}\}
/\frac{f^{(n)}(t_0)}{n!}
\end{equation}
 satisfies $\lim_{m\rightarrow
\infty}\mu_{m,n}(f;t_0,t_1) =1$. Expression (11) means that the
usual Taylor expansion at the point $t_1$ can be represented by the
so-called generalized Taylor expansion at the point $t_0$.  Let
$\xi_0$ be the nearest singular point from $t_1$ and
$h=\frac{\xi_0-t_0}{t_1-\xi_0}$, respectively, we have
$t_1=(\xi_0-t_0)(1+\frac{1}{h})+t_0$. Then the convergence region
$|t-t_1|<|t_1-\xi_0|$ becomes
\begin{equation}
|1+h-h\frac{t-t_0}{\xi_0-t_0}|<1.
\end{equation}
It is just the convergence region of another generalized Taylor
series given by Liao in Ref.[14] where the corresponding approaching
function is $\mu_{m,n}(h)=(-h)^{n}\sum_{k=0}^{m-n}
\binom{k+n-1}{k}(1+h)^{k}$ which doesn't depends on $f$. This means
that these two series have the same convergence region, and hence
our Taylor expansion is equivalent to the Liao's generalized Taylor
expansion. If $\xi_0$ is not a real number, let
$h=\frac{\eta_0-t_0}{t_1-\eta_0}$ respectively we have
$t_1=(\eta_0-t_0)(1+\frac{1}{h})+t_0$ where the real number $\eta_0$
satisfies $|t_1-\xi_0|=|t_1-\eta_0|$. Then the convergence region
$|t-t_1|<|t_1-\xi_0|$ becomes
\begin{equation}
|1+h-h\frac{t-t_0}{\eta_0-t_0}|<1.
\end{equation}
For example, we take  $f(t)=\frac{1}{1+t^2}$ and $t_0=0$. Therefore,
we have $\xi_0=\pm \mathrm{i}$ and $\eta_0=t_1\pm\sqrt{t_1^2+1}$. In
order that $t_0$ belongs to the convergence region of $f$ at the
point $t_1$, when $t_1>0$ or $t_1<0$, we have
$\eta_0=t_1-\sqrt{t_1^2+1}$ or $\eta_0=t_1+\sqrt{t_1^2+1}$
respectively.

\textbf{Remark}. According to different choice of the function $f$
and the points $t_0$ and $t_1$, we can construct infinite number of
approaching functions. For every approaching function, we can give a
kind of generalized Taylor expansion method. Therefore there exist
infinite number of generalized Taylor expansion methods. In his
book[13], Liao gives two kinds of generalized Taylor expansion
methods.

\section{On the series solution of the Blasius equation}

Blasius equation reads
\begin{equation}
f'''(\eta)+\frac{1}{2}f(\eta)f''(\eta)=0,
\end{equation}
with the conditions
\begin{equation}
f(0)=f'(0)=0, \ \ f'(+\infty)=1,
\end{equation}
which describes the two-dimensional viscous laminar flow over an
infinite flat-plain[15]. In 1908, Blasius [15] gave a series
solution
\begin{equation}
f(\eta)=\sum_{k=0}^{+\infty}(-\frac{1}{2})^k\frac{A_k\sigma^{k+1}}{(3k+2)!}\eta^{3k+2},
\end{equation}
where $\sigma=f''(0)$ and
\begin{equation}
A_0=A_1=1, \ \ A_k=\sum_{r=0}^{k-1}\binom{3k-1}{3r}A_rA_{k-r-1},
(2\leq k).
\end{equation}
Blasius obtained $\sigma\approx0.332$ and the convergence region
$0\leq\eta<\rho_0\approx5.690$. In 1997,  Liao obtained a
generalized Taylor series solution by means of the homotopy analysis
method,
\begin{equation}
f(\eta)=\lim_{m\rightarrow
\infty}\sum_{k=0}^{m}[(-\frac{1}{2})^k\frac{A_k\sigma^{k+1}}{(3k+2)!}\eta^{3k+2}]\mu_0^{m,k}(h),
\end{equation}
which converges in the region
\begin{equation}
-\rho_0\leq\eta\leq\rho_0[\frac{2}{|h|}-1]^{1/3},
\end{equation}
and where
\begin{equation}
\mu_0^{m,n}(h)=(-h)^n\sum_{k=0}^{m-n}\binom{n-1+k}{k}(1+h)^k.
\end{equation}

Now we use the Taylor expansion of $f(\eta)$ at the point $\eta_0$
to obtain the Liao's result. Assuming that
\begin{equation}
f(\eta)=\sum_{n=0}^{+\infty}a_n(\eta-\eta_0)^n,
\end{equation}
and substituting it into the Blasius equation, we have
\begin{equation}
a_{n+3}=-\frac{1}{2(n+3)(n+2)(n+1)}\sum_{m=0}^n(m+2)(m+1)a_{m+2}a_{n-m},
\ \  (n\ge0).
\end{equation}
In order to determine the values of $a_0, a_1$ and $a_2$, we use the
conditions (16) to give
\begin{equation}
\sum_{n=0}^{+\infty}a_n(-\eta_0)^n=0,
\end{equation}
\begin{equation}
\sum_{n=0}^{+\infty}(n+1)a_{n+1}(-\eta_0)^n=0,
\end{equation}
and
\begin{equation}
\lim_{\eta\rightarrow
+\infty}\sum_{n=0}^{+\infty}(n+1)a_{n+1}(\eta-\eta_0)^n=1,
\end{equation}
the condition (26) also can be replaced by using $\sigma$
\begin{equation}
\sum_{n=0}^{+\infty}(n+2)(n+1)a_{n+2}(-\eta_0)^n=\sigma\approx0.332.
\end{equation}
Then we can obtain the values of $a_n$ for $n=0,1,\cdots$.

If we take
\begin{equation}
\eta_0=\rho_0[\frac{2}{|h|}-1]^{1/3}\frac{1+h}{2+h},
\end{equation}
we find that the Liao's generalized Taylor series solution is just
the usual Taylor series solution at the point $\eta_0$. In fact, the
convergence region of our series solution is
$\eta_0-|\eta_0-\varsigma_0|<\eta<\eta_0+|\eta_0-\varsigma_0|$,
where the real number $\varsigma_0$ satisfies
$|\eta_0-\varsigma_0|=|\eta_0-\xi_0|$,  $\xi_0$ is the nearest
singular point from $\eta_0$. Of course, we don't know the value of
$\xi_0$. Therefore, our computation is an approximation treatment.
Since $\eta=0$  belongs to the region, we require $\varsigma_0<0$.
Furthermore, let
\begin{equation}
\eta_0=\varsigma_0(1+\frac{1}{h}),
\end{equation}
and $\eta_0>0$. By assuming that the convergence regions of two
series solutions are the same one, we take
$\eta_0+|\eta_0-\varsigma_0|=\rho_0[\frac{2}{|h|}-1]^{1/3}$ to give
\begin{equation}
\frac{(h+2)\eta_0}{h+1}=\rho_0[\frac{2}{|h|}-1]^{1/3},
\end{equation}
from which we obtain the result (28).

For example, we take $h=-\frac{1}{2}$. Correspondingly, we have
$\eta_0\approx 2.735$. The numerical result is coincident with
Liao's[3].

\section{Conclusion}
Through detailed analysis of some examples, we show that the
generalized Taylor series is only the usual Taylor expansion at
another point. This means that we can use the series expansion at
other point to give the same result obtained by the homotopy
analysis method. Our results uncover the essence of the generalized
Taylor expansion as the key of the homotopy analysis method.

\textbf{Acknowledgments}. I would like to thank the referees for
their valuable suggestions.

\end{document}